\documentclass[10pt,conference]{IEEEtran}

\IEEEoverridecommandlockouts
\usepackage{amsmath,amssymb,amsfonts}
\usepackage{algorithmic}
\usepackage{cite}
\usepackage{graphicx}
\usepackage{textcomp}
\usepackage{xcolor}

\usepackage[bottom]{footmisc}
\usepackage{hyperref}
\usepackage{footnotebackref}
\usepackage{cleveref}
\usepackage{booktabs}
\usepackage{subcaption}
\usepackage{listings}
\usepackage{courier}
\usepackage{array}
\usepackage{verbatim}
\usepackage{caption}
\usepackage{cancel}
\usepackage{multirow}
\usepackage{stmaryrd}
\usepackage{enumitem}
\usepackage{framed}
\usepackage{xargs}

\renewenvironmentx{leftbar}[2][1=1pt, 2=5pt]%
{\def\FrameCommand{\vrule width #1 \hspace{#2}}\MakeFramed {\advance\hsize-\width \FrameRestore}}%
{\endMakeFramed}

{\begin{leftbar}\begin{quotation}}%
{\end{quotation}\end{leftbar}}

\def\BibTeX{{\rm B\kern-.05em{\sc i\kern-.025em b}\kern-.08em
    T\kern-.1667em\lower.7ex\hbox{E}\kern-.125emX}}

\newcommand{\mercurial}[0]{Baseline}
\newcommand{\completion}[0]{Autocompletion}
\newcommand{\edit}[0]{Edit}

\begin{document}

\title{Learning Autocompletion from Real-World Datasets}

\author{
    \IEEEauthorblockN{Gareth Ari Aye}
    \IEEEauthorblockA{
        \textit{Facebook Inc.}\\
        Menlo Park, U.S.A. \\
        gaa@fb.com
    }
    \and
    \IEEEauthorblockN{Seohyun Kim}
    \IEEEauthorblockA{
        \textit{Facebook Inc.}\\
        Menlo Park, U.S.A. \\
        skim131@fb.com
    }
    \and
    \IEEEauthorblockN{Hongyu Li}
    \IEEEauthorblockA{
        \textit{Facebook Inc.}\\
        Menlo Park, U.S.A. \\
        hongyul@fb.com
    }
}

\maketitle

\begin{abstract}
Code completion is a popular software development tool integrated into all major IDEs. Many neural language models have achieved promising results in completion suggestion prediction on synthetic benchmarks. 
However, a recent study \textit{When Code Completion Fails: a Case Study on Real-World Completions} demonstrates that these results may not translate to improvements in real-world performance. 
To combat this effect, we train models on real-world code completion examples and find that these models outperform models trained on committed source code and working version snapshots by 12.8\% and 13.8\% accuracy respectively. 
We observe this improvement across modeling technologies and show through A/B testing that it corresponds to a 6.2\% increase in programmers' actual autocompletion usage. 
Furthermore, our study characterizes a large corpus of logged autocompletion usages to investigate why training on real-world examples leads to stronger models.
\end{abstract}


\section{Introduction}
\begin{figure}[]
\includegraphics[width=\columnwidth]{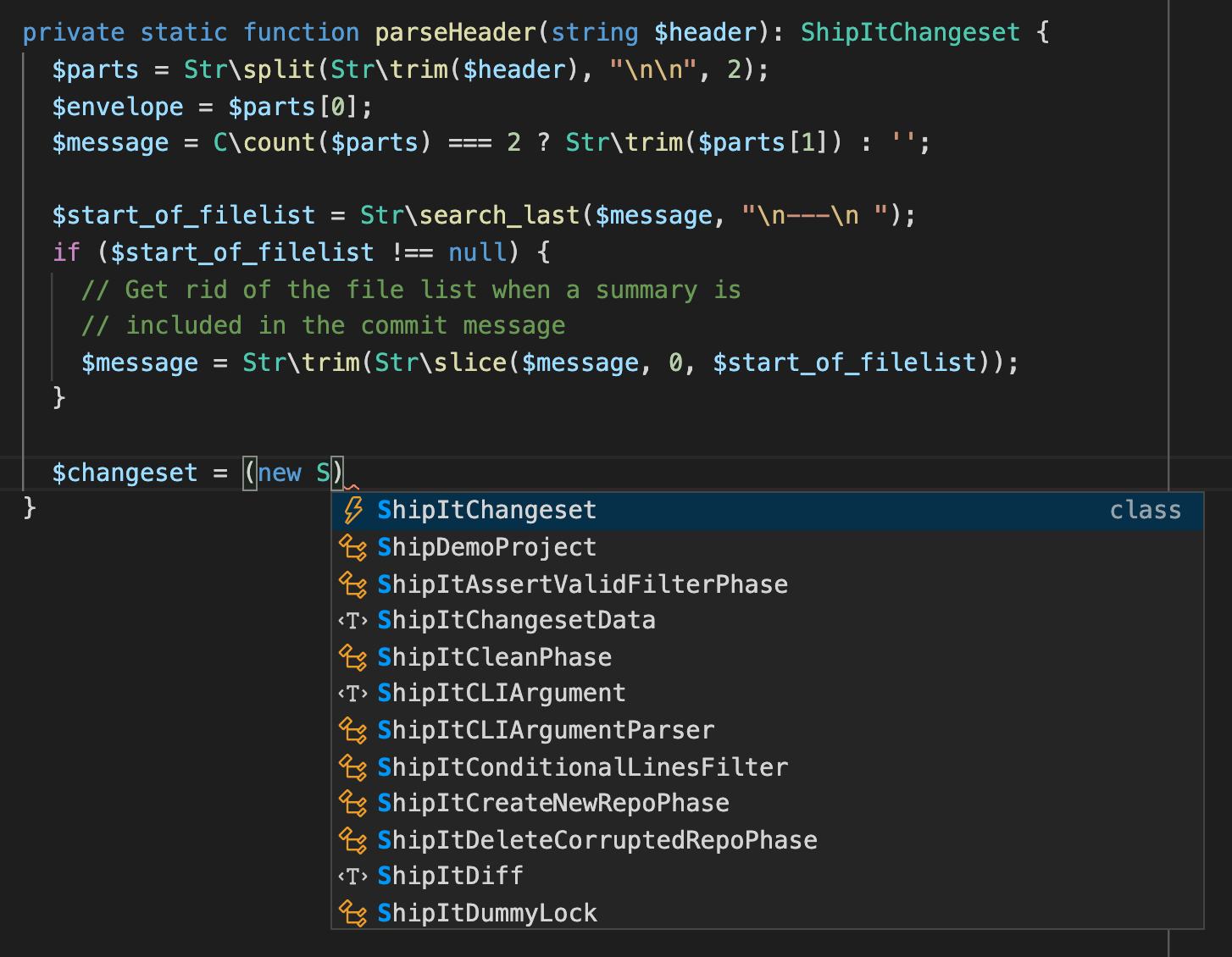}
\caption{Example of IDE autocompletion\protect\footnotemark}
\label{fig:ide}
\end{figure}

Autocompletion is the most frequently used IDE feature\cite{Murphy:2006:JSD:1159169.1159396}. Beyond its utility as a typing assistant, code completion also helps software developers discover contextually relevant APIs and libraries. Fig~\ref{fig:ide} shows an example of IDE autocompletion. Significant attention has been given to improving suggestion ranking through machine learning\cite{Bruch:2009:LEI:1595696.1595728,Hindle:2012:NS:2337223.2337322,aye2020sequence,kim2020code} by feeding code to models as a sequence of tokens or even AST nodes\cite{Li_2018}. Recent neural code completion studies evaluate models by taking existing source code and measuring how accurately the probability $P(t_i | t_0, t_1, ..., t_{i-1})$ of tokens $t_i$ can be predicted given preceding context $t_0, t_1, ..., t_{i-1}$\cite{aye2020sequence,kim2020code,Li_2018,Raychev:2014:CCS:2594291.2594321,DBLP:journals/corr/abs-1903-05734}. 

However, existing literature focuses model training and evaluation on committed code as it appears version control. Raychev et al.\cite{10.1145/2983990.2984041} leveraged a collection of 150k committed Python files from GitHub. The benchmark used in Brockschmidt et al.\cite{brockschmidt2019generative} is composed of 593 popular, open-source C\# projects. Alon et al.\cite{alon2020structural} evaluated models on eleven large GitHub Java repositories. These training corpora lack samples of code that is undergoing active modification. Software development tools such as autocompletion make several important assumptions when integrating models trained on committed code:

\begin{enumerate}
\item{code is written sequentially from left to right,}
\item{code undergoing active modification is sufficiently similar to code that has already been written, and}
\item{correct completion suggestions follow the same distribution as tokens in existing source code.}
\end{enumerate}

\footnote{\url{https://github.com/facebook/fbshipit/blob/master/src/shipit/repo/ShipItRepoGIT.php\#L123}}

Hellendoorn et al.\cite{hellendoorn-codefails} show that these assumptions do not hold in practice and that models trained and evaluated on existing source code suffer from \textit{concept drift} when applied to real-world completion examples. Since the ultimate goal of research on improving software language models is to build better real-world software development tools, it is important to evaluate models in a way so that better benchmark performance corresponds to better performance in practice. We find that accuracy falls from 0.462 to 0.203 (see Section~\ref{sec:results}) when a state-of-the-art language model is evaluated on real-world examples (logged autocompletion events from IDE usage at Facebook) compared to evaluation on identifiers randomly sampled from code in version control. While training and evaluating models on committed code can determine the most powerful language models in the abstract, this approach is inaccurate for measuring improvements to real-world IDE code completion.

This study compares training and evaluating identical language models on existing source code against real-world examples. 
We answer the following research questions:

\begin{leftbar}
\noindent \textit{RQ1: Does software language model performance vary when real-world autocompletion events are used for evaluation instead of identifiers sampled from code in version control?}
\end{leftbar}

Our experimental results show that models trained on committed source code perform significantly worse when scored on real-world autocompletion events instead of random, held-out identifier samples from version control. Improvements in sequence modeling performance on committed code may not translate to real-world autocompletion improvements.\\

\begin{leftbar}
\noindent\textit{RQ2: Do language models trained on real-world examples better predict developers’ completion suggestion selections?}
\end{leftbar}

We show that prediction accuracy improves significantly across neural and count-based language models when learning from real-world examples instead of code in version control.\\

\begin{leftbar}
\noindent\textit{RQ3: When deployed in the IDE to rank completion suggestions, do language models trained on real-world examples drive greater tool usage?}
\end{leftbar}

We prove through A/B testing with thousands of software developers that autocompletion usage increases with statistical signficance when a model trained on real-world examples ranks completion suggestions in the IDE.\\

\begin{leftbar}
\noindent\textit{RQ4: What are the differences between code in source control and autocompletion that result in divergent completion ranking behavior?}
\end{leftbar}

Compared to code in source control, code undergoing active modification in the IDE is much less likely to parse and compile at any given point in time. Additionally, many idioms related to debugging, logging, and ad hoc testing are much more common in IDE code than in source control. Tokens from accepted completion suggestions are longer than average identifiers in committed code. Compared to autocompletion events, code in version control may also contain examples that have lost relevance over time as new code is written and old code is deprecated. However, performance does not improve when training models on either working version snapshots or a limited subset of recently-edited files in version control, leading to the conclusion that recency bias in real-world examples is not a major factor.\\

\noindent\textbf{Contributions}
\begin{enumerate}
    \item We train and evaluate a variety of software language models on real and artificial autocompletion examples. When evaluating on a real-world benchmark sourced from logged developer activity, models trained on real code completion examples outperform models trained on committed code. 
    \item We describe and share results from several A/B tests conducted with thousands of programmers at Facebook to show that models trained on logged code completion examples drive greater real-world tool usage.
    \item We explore and discuss differences between random identifiers appearing in source control and tokens applied in code completion that make the former an insufficient benchmark for modeling the latter.
\end{enumerate}

\vspace{2mm}

\noindent\textbf{Outline}

\vspace{1mm}

The rest of this paper is organized to first introduce our experimental setup in \Cref{sec:experiments}. In this section we describe the corpora, language models, and evaluation methodologies employed in our study. \Cref{sec:results} reports experimental results, supporting the answers to our research questions. \Cref{sec:threats}, \Cref{sec:relatedwork}, and \Cref{sec:future} discuss threats to validity, related work, and future work, respectively. Finally, \Cref{sec:conclusion} concludes with a summation and key insights for autocomplete tool designers.

\section{Experimental Setup}
\label{sec:experiments}

\subsection{Datasets} \label{sec:datasets}

Software language models are typically trained and evaluated offline on large corpora of existing source code\cite{Bruch:2009:LEI:1595696.1595728,Hindle:2012:NS:2337223.2337322,aye2020sequence,kim2020code,Raychev:2014:CCS:2594291.2594321,DBLP:journals/corr/abs-1903-05734}. Our approach examines the performance of several identical language models trained on samples from three datasets summarized in Table I.

\begin{table*}[]
\centering
\caption{Summary of code sequence datasets}
\def\arraystretch{1.20}
\begin{tabular}{l|l|l|l|l}
\toprule
Dataset & Type & Description & Size & \# of tokens \\
\midrule
\mercurial{} & Synthetic & Source files randomly sampled from version control repository & 977k files & 333,115,103 \\
\completion{} & Real & Accepted completion suggestions logged in the IDE & 3.4m completion acceptances & 257,217,900 \\
\edit{} & Mixed & Code edits logged from each file-save operation in the IDE & 1.6m edit sequences & 161,590,004 \\
\bottomrule
\end{tabular}
\label{table:data-used}
\end{table*}

\begin{itemize}
    \item \textbf{\mercurial{}}: Nearly one million source files containing over three hundred million tokens written in the Hack dialect of PHP. These source files are collected from Facebook's internal source code repository. This dataset is similar to the committed, open-source code on GitHub used in existing literature. It contains already-written code and assumes that tokens used in autocompletion follow the same distribution as identifiers in version control source code. 
    \item \textbf{\completion{}}: Over three million code completion acceptance events logged over the course of several months capturing applied suggestions with surrounding context from Hack programmers' autocompletion activity. This dataset consists of real-world examples from actual developer activity within the IDE. 
    \item \textbf{\edit{}}: Nearly two million snapshots captured over the course of several months containing working versions of source files modified by Hack programmers. Although this dataset contains code undergoing active modification, only a small subset of each snapshot will typically be modified. As in \textit{\mercurial{}}, identifiers are randomly sampled from these snapshots.
\end{itemize}

Notably code in the \textit{\mercurial{}} corpus passes through a peer-review process and is guaranteed to parse and compile. Conversely, samples from \textit{\completion{}} and \textit{\edit{}} capture code as it is being written before any checks or reviews. \textit{\mercurial{}} follows the standard approach where training and evaluation token sequences are randomly sampled from a corpus of already-written code. By contrast, \textit{\completion{}} contains samples directly sourced from real developer activity in the IDE. The purpose of the third working version snapshots dataset \textit{\edit{}} is to validate that improvements in model accuracy cannot simply be attributed to the fact that \textit{\completion{}} biases toward fresh, recently-edited code. \textit{\edit{}} shares this bias, so we would expect \textit{\edit{}} to outperform \textit{\mercurial{}} if recency bias is an important contributing factor to the improved performance of models trained on \textit{\completion{}}.

\subsection{Models}

To show that our findings generalize across modeling technologies, we test several different language models for next token prediction. Since our study focuses on the impact of selecting different training and evaluation corpora, we do not introduce novel model configurations or compare against other state-of-the-art models from prior literature. Our experiments leverage n-gram, whole token transformer, and subtoken transformer models. Each of the three models is briefly described below. Since the focus of this paper is not to evaluate the power of different models, we do not go into detailed descriptions of the models.

\subsubsection{N-gram model}
N-gram models predict the probability $p(t_n | t_{1}, t_{2}, ..., t_{n-1})$ by directly counting the percentage of training samples in which token $t_n$ follows the subsequence $t_{1}, t_{2}, ..., t_{n-1}$. Owing to their relative simplicity, n-gram models are commonly used in sequence modeling. While model training and prediction is fast for an n-gram model, one disadvantage is the inability to handle unseen or infrequent n-grams. To overcome this limitation, we use the KenLM\footnotemark{}\cite{heafield-2011-kenlm, heafield-etal-2013-scalable} library in our work, which incorporates n-gram language modeling using modified Kneser-Ney smoothing\cite{10.3115/981863.981904} without pruning. 

\footnotetext{\url{https://github.com/kpu/kenlm}}

\subsubsection{Transformer}
\begin{figure}[]
\includegraphics[width=\columnwidth]{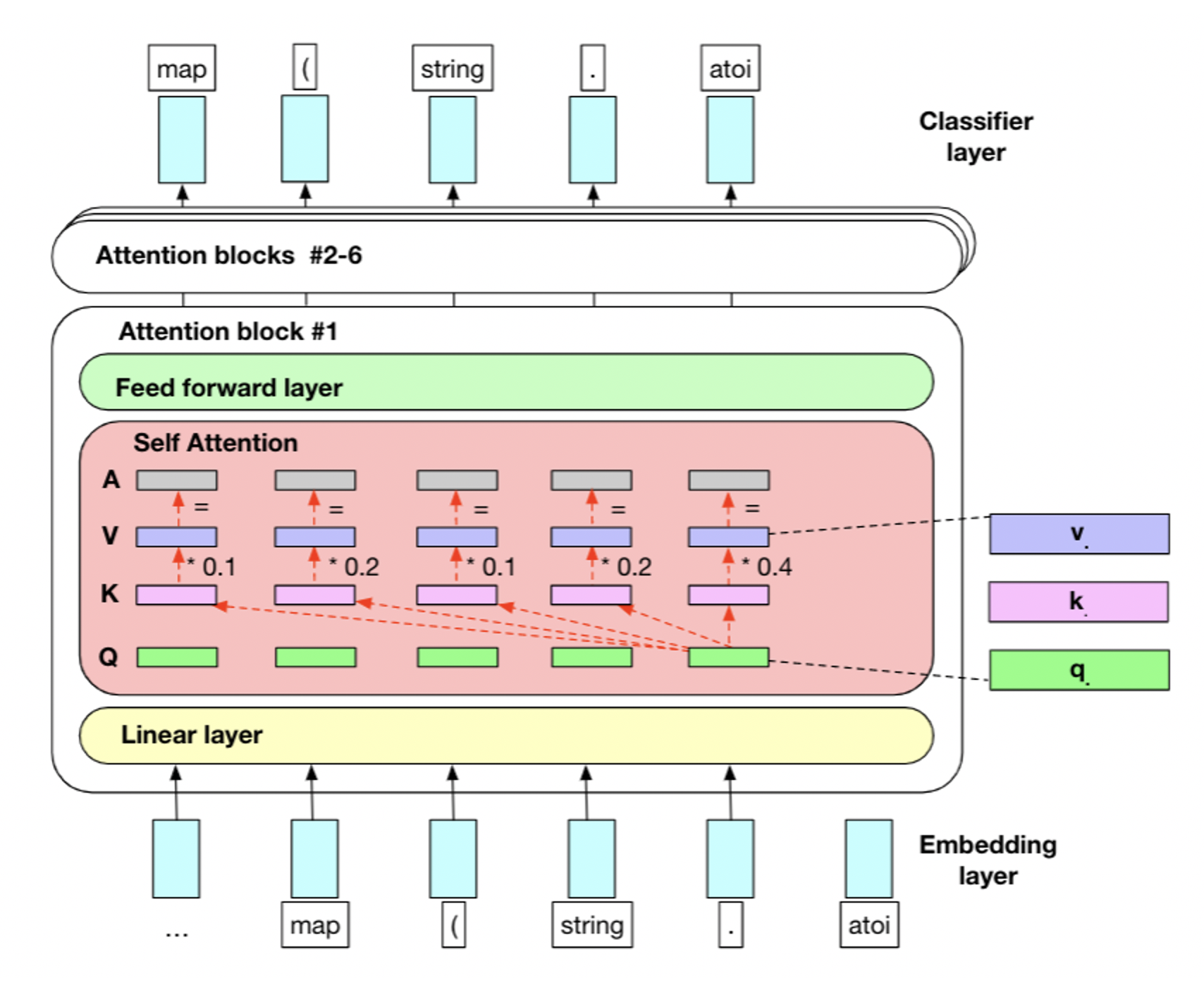}
\caption{Architecture of the GPT-2 Transformer model}
\label{fig:gpt2}
\end{figure}

Transformers are deep neural networks that have achieved state-of-the-art performance in sequence modeling, surpassing previous model architectures such as LSTM and 1D-CNN. In language modeling, Transformers have the advantage of observing all sequence elements simultaneously, allowing them to easily build long-distance connections using self-attention. With self-attention, importance weights are computed between each pair of tokens in the input sequence. This is especially effective in code prediction, as non-local variable and function definitions can have high relevance. Fig ~\ref{fig:gpt2} shows an overview of the Transformer model architecture (we use the GPT-2 implementation from \cite{kim2020code})\footnote{Implementation of the GPT-2 model is available at \url{https://github.com/facebookresearch/code-prediction-transformer/blob/master/model.py}}. At a high-level, sequences are processed through an embedding layer, then through six self-attention blocks, and finally a softmax classification layer, resulting in a probability distribution over next token candidates. 

\subsubsection{Byte pair encoding}
One difficulty in modeling source code as compared to natural language is that code introduces new vocabulary at a far higher rate\cite{DBLP:journals/corr/abs-1903-05734}. Recognizing and predicting rare tokens from large, open-source code vocabularies poses a challenge. One solution is to model code as a sequence of partial (rather than whole) tokens; byte pair encoding (BPE)~\cite{bpe} is a popular technique for building subtoken vocabularies. BPE learns to tokenize words as subword units based on the frequency of certain character combinations. This method is promising for source code tokenization, as rare names sharing similar contexts are often composed of several more common partial names. In our evaluation, we observe significant improvements in accuracy by replacing whole-token encoding with BPE.

\subsection{Training}

Each of the three datasets listed in \cref{sec:datasets} is separated into training, validation, and test samples in an 8:1:1 ratio. \textit{\mercurial{}} $\cup$ \textit{\completion{}} is treated as an additional training corpus in order to explore the effects of combining real and artificial data in model training. Thus, each language model is trained on four different corpora:
\begin{enumerate}
    \item \textit{\mercurial{}}
    \item \textit{\completion{}}
    \item \textit{\edit{}}
    \item \textit{\mercurial{}} $\cup$ \textit{\completion{}}
\end{enumerate}

During data preprocessing, we select a distinct vocabulary for each training corpus (\textit{\mercurial{}}, \textit{\completion{}}, \textit{\edit{}}, and \textit{\mercurial{} $\cup$ \completion{}}). Vocabularies are computed by aggregating all corpus tokens and choosing the top 100k and 10k most frequent names for whole token and BPE encodings respectively. 
The vocabulary sizes cover more than 99\% of the tokens in each of the training corpora.
For the n-gram and whole-token Transformer models, 
training samples from each dataset are composed of sequences of up to 100 (context length) consecutive tokens $t_1, t_2, ..., t_{100}$. 
For the Transformer model using BPE, since partial tokenization will increase the sequence length, we extract a token sequence (of up to length 100) before applying BPE encoding. We increase the context length to 300 so that the model can take the same amount of context information.

The training objective for each language model is to estimate the conditional probability $p(t_i | t_1, t_2, ..., t_{i-1})$. Out-of-vocabulary (OOV) tokens are replaced with a special \lstinline{<unk>} token and sequences shorter than 100 tokens (or 300 for Transformer + BPE) are right-padded with \lstinline{<pad>}.

We train Transformer models on each of the training corpora for a maximum of 15 epochs with early stopping\footnote{Patience value 2 is used for early stopping where training is terminated if no improvement on validation data is observed for two consecutive epochs.} using Nvidia Tesla V100 GPUs.

\subsection{Offline Evaluation}

After training models on samples from each of the four datasets, we evaluate performance on the held-out test samples from \textit{\mercurial{}} and \textit{\completion{}}. For \textit{\mercurial{}} we randomly select identifier tokens (e.g. local variable, method invocation, etc.) from the test sequences for evaluation. In the case of \textit{\completion{}}, the models are specifically evaluated on completion tokens accepted by real IDE users. We use two standard metrics for evaluation: top-1 accuracy and mean reciprocal rank (MRR). Top-1 accuracy simply measures in what percentage of test sequences $t_1, t_2, ..., t_k$ a model correctly predicts that $t_k$ follows the subsequence $t_1, t_2, ..., t_{k-1}$. MRR gives partial credit equal to the inverse rank that a model assigns the correct token (e.g. 0.5 if a model predicts token $t_k$ in two guesses).
It is defined as:  
\begin{equation}
\text{MRR} = \frac{1}{n}\sum_{i=1}^{n}\frac{1}{rank_{i}}
\end{equation}
where $n$ is the number of testing datapoints, and $rank_i$ is the rank of the correct token predicted by the model as the i\textsuperscript{th} candidate. In our evaluation, we only consider the top ten results (otherwise the score will be zero) as showing additional suggestions is impractical in the IDE context.

\subsection{A/B Testing}

\begin{figure*}[]
\includegraphics[width=2\columnwidth]{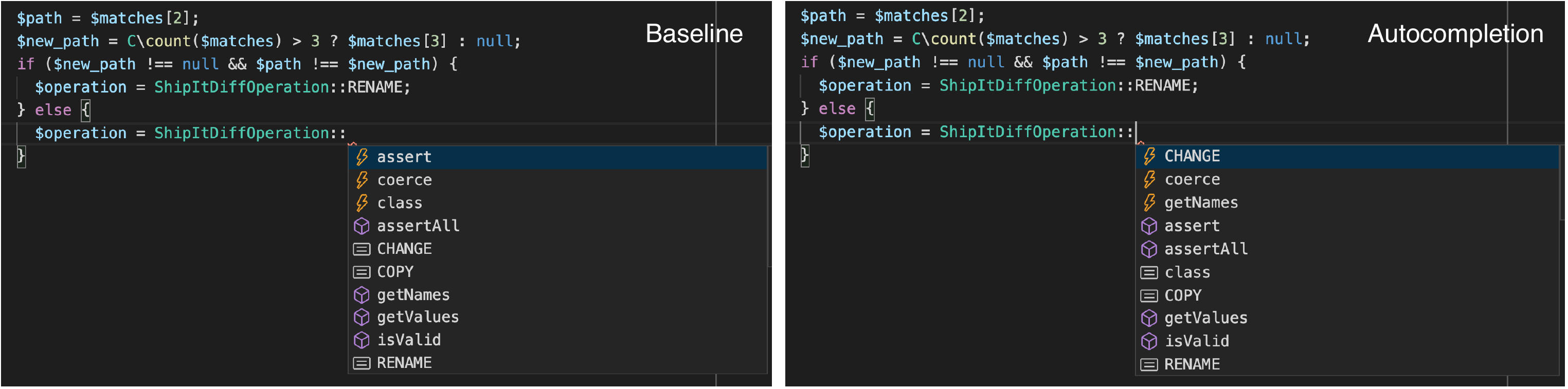}
\centering
\caption{Comparison of autocomplete results\protect\footnotemark{} ranked by Transformer models trained on \textit{\mercurial{}} (left) and \textit{\completion{}} corpus (right) in the IDE.}
\label{fig:live-example}
\end{figure*}

In order to answer \textit{RQ3} regarding whether or not improvements in model accuracy lead to greater autocompletion utility for end users, we perform two live A/B tests on thousands of software developers internally. We integrate autocompletion ranking into the IDE for these developers so that up to three high probability completion suggestions can be prioritized by descending model probability and displayed at the top of the completion dropdown window (see \cref{fig:ide} and \cref{fig:live-example}), with remaining completion items ordered alphabetically. A threshold $t = 0.1$ is applied where only suggestions $x$ with model probability $p(x) > t$ are prioritized so that low confidence predictions can be omitted.

\footnotetext{\url{https://github.com/facebook/fbshipit/blob/master/src/shipit/repo/ShipItRepo.php\#L161}}

In each test, developers are randomly assigned to one of three groups corresponding to a model trained on one of \textit{\mercurial{}}, \textit{\completion{}}, or \textit{\edit{}}. These experiments treat \textit{\mercurial{}} as the control group and models trained on other datasets as experiment groups. We frame the number of completion suggestions accepted by a given developer on a given day as an observation and record autocompletion acceptance events in order to compute population statistics for each experiment group. Both experiments are conducted until reaching statistical significance at or above a 90\% confidence level for the relationship between population means.

\section{Analysis}\label{sec:results}

\begin{leftbar}
\noindent \textit{RQ1: Does software language model performance vary when real-world autocompletion events are used for evaluation instead of identifiers sampled from code in version control?}
\end{leftbar}

Yes. Accuracy in next token prediction on code in version control does not translate to real-world autocompletion events. From Table~\ref{tab:results}, we see that top-1 accuracy of an n-gram model trained on \textit{\mercurial{}} drops from 0.183 to 0.081 when evaluated on \textit{\completion{}} instead of held-out \textit{\mercurial{}} examples. 
The Transformer model also sees a reduction of over 50\% in top-1 accuracy (0.462 to 0.203). For the Transformer + BPE model, the drop in accuracy is relatively lower compared to n-gram and Transformer (0.542 to 0.312). Nonetheless, there is still a significant decrease. We observe similar trends in MRR evaluation (\cref{tab:results-mrr}).

These findings demonstrate that evaluating autocompletion models on version control datasets does not indicate how well they will perform in the IDE setting. In RQ2, we address how to bridge the gap between the two evaluation datasets. RQ3 discusses to what extent evaluation on the \textit{\completion{}} dataset corresponds to improvements in developer experience. Finally, RQ4 tries to examine what makes the two evaluation datasets different.

\begin{table}[]
	\caption{Summary of offline evaluation results. Each model is trained on the four training corpora and evaluated on held-out \textit{\mercurial{}} and \textit{\completion{}} text examples.}
	\def\arraystretch{1.25}

\begin{subtable}{\columnwidth}
\centering
\subcaption{Top-1 Accuracy}
\vspace{-1mm}
\begin{tabular}{c|c|c|c}
\toprule
\multirow{2}{*}{Model} & \multirow{2}{*}{Training Corpus} & \multicolumn{2}{c}{Evaluation Corpus} \\
\cline{3-4}
 & & \mercurial{} & \completion{} \\
\midrule
\multirow{5}{*}{n-gram} & \mercurial{} & \textbf{0.183} & 0.081 \\
\cline{2-4}
 & \completion{} & 0.144 & \textbf{0.144} \\
\cline{2-4}
 & \edit{} & 0.157 & 0.085 \\
\cline{2-4}
 & \mercurial{} $\cup$ & \multirow{2}{*}{0.180} & \multirow{2}{*}{0.126} \\
 & \completion{} & & \\
\midrule
\multirow{5}{*}{Transformer} & \mercurial{} & \textbf{0.462} & 0.203 \\
\cline{2-4}
 & \completion{} & 0.319 & \textbf{0.331} \\
\cline{2-4}
 & \edit{} & 0.323 & 0.193 \\
\cline{2-4}
 & \mercurial{} $\cup$ & \multirow{2}{*}{0.457} & \multirow{2}{*}{0.319} \\
 & \completion{} & & \\
\midrule
\multirow{5}{*}{Transformer + BPE} & \mercurial{} & \textbf{0.542} & 0.312 \\
\cline{2-4}
 & \completion{} & 0.437 & 0.353 \\
\cline{2-4}
 & \edit{} & 0.432 & 0.284 \\
\cline{2-4}
 & \mercurial{} $\cup$ & \multirow{2}{*}{0.532} & \multirow{2}{*}{\textbf{0.362}} \\
 & \completion{} & & \\
\bottomrule
\end{tabular}
\label{tab:results-top1}
\end{subtable}

\vspace{+3mm}

\begin{subtable}{\columnwidth}
\centering
\subcaption{MRR}
\vspace{-1mm}
\begin{tabular}{c|c|c|c}
\toprule
\multirow{2}{*}{Model} & \multirow{2}{*}{Training Corpus} & \multicolumn{2}{c}{Evaluation Corpus} \\
\cline{3-4}
 & & \mercurial{} & \completion{} \\
\midrule
\multirow{5}{*}{n-gram} & \mercurial{} & \textbf{0.282} & 0.145 \\
\cline{2-4}
 & \completion{} & 0.244 & \textbf{0.250} \\
\cline{2-4}
 & \edit{} & 0.230 & 0.155 \\
\cline{2-4}
 & \mercurial{} $\cup$ & \multirow{2}{*}{0.283} & \multirow{2}{*}{0.223} \\
 & \completion{} & & \\
\midrule
\multirow{5}{*}{Transformer} & \mercurial{} & \textbf{0.524} & 0.267 \\
\cline{2-4}
 & \completion{} & 0.383 & \textbf{0.408} \\
\cline{2-4}
 & \edit{} & 0.381 & 0.247 \\
\cline{2-4}
 & \mercurial{} $\cup$ & \multirow{2}{*}{0.519} & \multirow{2}{*}{0.396} \\
 & \completion{} & & \\
\midrule
\multirow{5}{*}{Transformer + BPE} & \mercurial{} & \textbf{0.615} & 0.399 \\
\cline{2-4}
 & \completion{} & 0.513 & \textbf{0.461} \\
\cline{2-4}
 & \edit{} & 0.517 & 0.368 \\
\cline{2-4}
 & \mercurial{} $\cup$ & \multirow{2}{*}{0.606} & \multirow{2}{*}{0.459} \\
 & \completion{} & & \\
\bottomrule
\end{tabular}
\label{tab:results-mrr}
\end{subtable}

\label{tab:results}
\end{table}

\begin{leftbar}
\noindent\textit{RQ 2: Do language models trained on real-world examples better predict developers' completion suggestion selections?}
\end{leftbar}

Yes. 
In our offline evaluation, we observe an improvement from 0.081 to 0.144 top-1 accuracy when training the n-gram model on \textit{\completion{}} instead of \textit{\mercurial{}} (see Table~\ref{tab:results}). There is an even larger gain for the Transformer model from 0.203 to 0.331. Although Transformer + BPE generalizes better from \textit{\mercurial{}} to \textit{\completion{}} examples, we still observe a major improvement from 0.312 to 0.353 top-1 accuracy from training on \textit{\completion{}}.

Furthermore, taking the union of different training corpora does not drive improvements in accuracy or MRR. In other words, the benefit of more training data is overshadowed by the drawback of concept drift. Training Transformer on \textit{\completion{}} (without sourcing additional examples from \textit{\mercurial{}} or \textit{\edit{}}) results in the best overall performance. For the Transformer + BPE model, \textit{\mercurial{}} $\cup$ \textit{\completion{}} performs only slightly better than using \textit{\completion{}} by itself despite tripling the number of training examples.

Evaluating on the \textit{\mercurial{}} dataset tells a different story. Among all of the datasets, training on the \textit{\mercurial{}} dataset yields the best performance on held-out examples from \textit{\mercurial{}}. At a high-level, we interpret that drawing software language model training examples from the same corpus that test examples are sourced from leads to the highest model accuracy. It is interesting to note that training on the \textit{\completion{}} dataset does a better job of generalizing when evaluated on another corpus. When evaluated on the other corpus, \textit{\completion{}} accuracy decreases 0.012 (0.331 to 0.319) while \textit{\mercurial{}} decreases 0.259 (0.462 to 0.203). 
For ranking real-world autocompletion suggestions, this insight suggests leveraging historical autocompletion usage for model training.\\

\begin{table*}[]
\caption{Production A/B test results. An observation counts the number of completion suggestions accepted by a given developer on a given day.}
\centering
\def\arraystretch{1.20}
\small
\begin{tabular}{c|c|c|c|c|c|c|c}
\toprule
Model & Training Corpus & observations & mean & std dev & \# unique developers & improvement & p-value \\
\midrule
\multirow{3}{*}{n-gram} & \mercurial{} & 3578 & 17.281 & 24.893 & 1857 & - & - \\
\cline{2-8}
 & \completion{} & 3711 & 18.272 & 25.626 & 1903 & 0.057 & 0.094 \\
\cline{2-8}
 & \edit{} & 3345 & 17.333 & 24.571 & 1774 & 0.003 & 0.931 \\
\midrule
\multirow{3}{*}{Transformer} & \mercurial{} & 4711 & 18.096 & 26.329 & 2048 & - & - \\
\cline{2-8}
 & \completion{} & 4848 & 19.227 & 27.373 & 2093 & 0.062 & 0.040 \\
\cline{2-8}
 & \edit{} & 4421 & 18.225 & 26.530 & 1930 & 0.007 & 0.816 \\
\bottomrule
\end{tabular}
\label{tab:ab-results}
\end{table*}

\begin{leftbar}
\noindent\textit{RQ3: When deployed in the IDE to rank completion suggestions, do language models trained on real-world examples drive greater tool usage?}
\end{leftbar}

Yes. To measure the extent to which our offline results correspond to improvements in developer experience, we conduct A/B tests where models trained on different corpora are used to rank code completion suggestions from the IDE. \Cref{tab:ab-results} shows the results. 

Programmers whose completion suggestions are ranked by an n-gram model trained on \textit{\completion{}} accept 5.7\% more suggestions per day than programmers receiving completion suggestions ranked by an n-gram model trained on \textit{\mercurial{}}. There is an even larger 6.2\% improvement when \textit{\completion{}} and \textit{\mercurial{}} Transformer models rank completion suggestions for the two groups' programmers. Both improvements are statistically significant with p-values of 0.094 and 0.04 respectively.

On the other hand, software developers whose completion suggestions are ranked by models trained on \textit{\edit{}} do not accept more daily suggestions compared to \textit{\mercurial{}}. The population means are equal, with statistical significance at p-values of 0.069 and 0.184 for the n-gram and Transformer experiments respectively.


These results, combined with our findings from offline testing, provide compelling evidence that training on the \textit{\completion{}} dataset yields the best real-world performance and should be deployed to drive greater tool usage.

\begin{figure}[]
\centering
\includegraphics[width=\columnwidth]{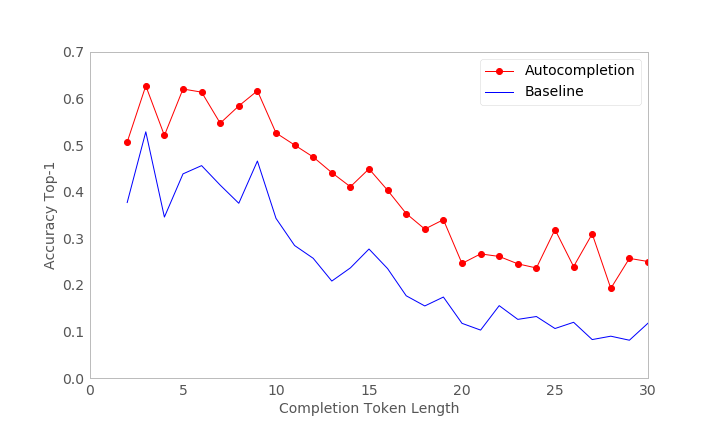}
\caption{Length of target tokens vs top-1 accuracy for \textit{\mercurial{}} and \textit{\completion{}} models evaluated on held-out \textit{\completion{}} examples. Both models have a harder time predicting longer tokens, but \textit{\completion{}} training model performs consistently better.}
\label{fig:accuracy-len}
\end{figure}

\begin{figure}[]
\centering
\includegraphics[width=\columnwidth]{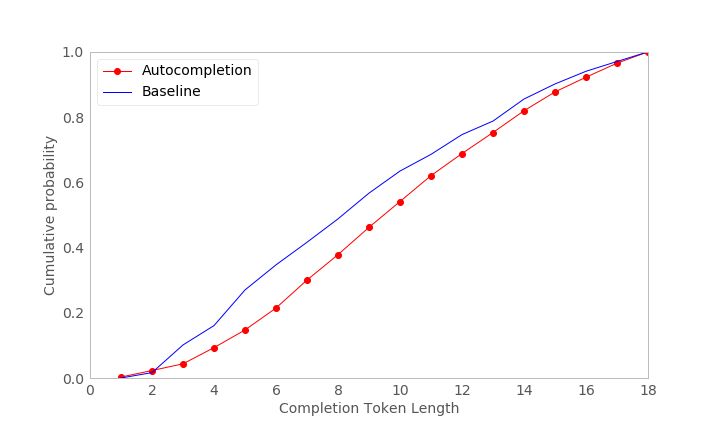}
\caption{Cumulative distribution function (CDF) of target tokens lengths for \textit{\mercurial{}} and \textit{\completion{}} training examples. Although short identifiers appear frequently in version control, they are seldom used in code completion.}
\label{fig:accuracy-len-cdf}
\end{figure}

\begin{table}[]
	\caption{OOV rates for \textit{Transformer} and \textit{Transformer + BPE} models on held-out test examples from the \textit{\mercurial{}} and \textit{\completion{}} datasets. The \textit{Transformer} vocabulary consists of 100k tokens whereas the \textit{Transformer + BPE} vocabulary consists of 10k partial tokens.}
	\def\arraystretch{1.20}
	\small

\begin{subtable}{\columnwidth}
\centering
\subcaption{OOV rates for the target tokens (token at the point of prediction)}
\vspace{-1mm}
\begin{tabular}{c|c|c|c}
\toprule
\multirow{2}{*}{Model} & \multirow{2}{*}{Training Corpus} & \multicolumn{2}{c}{Evaluation Corpus}  \\
\cline{3-4}
 & & \mercurial{} & \completion{} \\
\midrule
\multirow{2}{*}{Transformer} & \mercurial{} & 27.04\% & 33.85\% \\
\cline{2-4}
 & \completion{} & 32.96\% & 25.83\%  \\
\midrule
Transformer + & \mercurial{} & 0.25\% & 0.29\% \\
\cline{2-4}
BPE & \completion{} & 0.31\% & 0.31\% \\
\bottomrule
\end{tabular}
\label{tab:oov}
\end{subtable}

\vspace{+5mm}

\begin{subtable}{\columnwidth}
\centering
\subcaption{OOV rates for the context tokens (tokens that come before the point of prediction)}
\vspace{-1mm}
\begin{tabular}{c|c|c|c}
\toprule
\multirow{2}{*}{Model} & \multirow{2}{*}{Training Corpus} & \multicolumn{2}{c}{Evaluation Corpus}  \\
\cline{3-4}
 & & \mercurial{} & \completion{} \\
\midrule
\multirow{2}{*}{Transformer} & \mercurial{} & 21.72\% & 25.00\% \\
\cline{2-4}
 & \completion{} & 27.44\% & 18.90\%  \\
\midrule
Transformer + & \mercurial{} & 0.25\% & 0.23\% \\
\cline{2-4}
BPE & \completion{} & 0.26\% & 0.23\% \\
\bottomrule
\end{tabular}
\label{tab:oov-context}
\end{subtable}

\end{table}

\begin{leftbar}
\noindent\textit{RQ4: What are the differences between code in source control and autocompletion that result in divergent completion ranking behavior?}
\end{leftbar}

To answer this question, we compare and contrast five characteristics of these datasets: the out-of-vocabulary (OOV) rate for target tokens, OOV rate for context tokens, target token lengths, target token kinds, and recency of example sequences (e.g. time since last modification).

\paragraph{Target token OOV rate}
For the Transformer model, the vocabulary size is set to 100k tokens. While this covers over 99\% of the tokens in each of the training datasets\footnote{Respectively 18.66\% and 21.04\% of identifiers in \textit{\mercurial{}} and \textit{\completion{}} belong to their corresponding vocabulary containing the 100k most common names.} (including punctuation and keywords),
the rate of OOV target tokens increases dramatically when non-matching corpora are used for training and evaluation. For example, in the \textit{\mercurial{}} training corpus, the OOV rate increases from 27.04\% to 33.85\% when evaluated on \textit{\completion{}} rather than \textit{\mercurial{}} sequences (see Table~\ref{tab:oov}). 

For the Transformer + BPE model, the vocabulary size is set to 10k. Table~\ref{tab:oov} highlights the usefulness of subtoken encoding schemes like BPE, as the OOV rate dramatically falls to below 1\% across all evaluation corpora. Using BPE, Transformer models can learn to accurately embed rare or even novel tokens. This capability results in a lower relative drop in accuracy when Transformer + BPE is trained on one dataset and evaluated on another (see Table~\ref{tab:results}). This supports the finding in Karampatsis et al.\cite{DBLP:journals/corr/abs-1903-05734} that subtoken encoding can be especially helpful when developer activity datasets are unavailable for model training.

\begin{figure}[]
\includegraphics[width=\columnwidth]{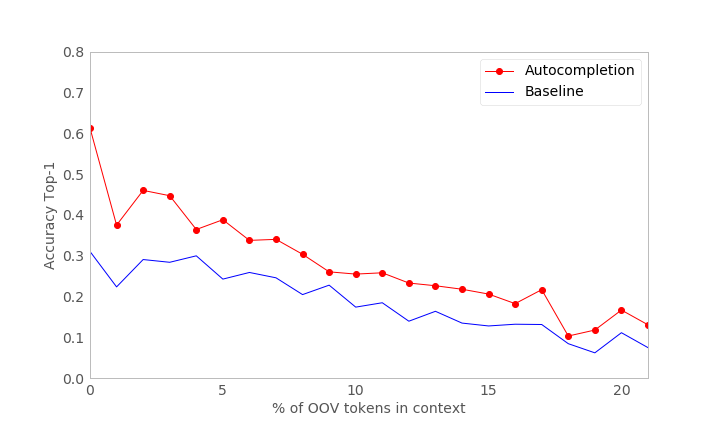}
\caption{Percentage of OOV tokens in the context vs top-1 accuracy for \textit{\mercurial{}} and \textit{\completion{}} Transformer models. There is an inverse correlation between OOV tokens in the context query and accuracy when these models are evaluated on the \textit{\completion{}} dataset.}
\label{fig:accuracy-oov}
\end{figure}

\paragraph{Context token OOV rate}
We also explore the impact of OOV tokens in the context tokens $t_0, t_1, ..., t_{n-1}$ (used to predict next token $t_n$) that appear before the cursor. As expected, we observe an inverse correlation between the number of OOV tokens and accuracy in both \textit{\mercurial{}} and \textit{\completion{}} Transformer models (\cref{fig:accuracy-oov}). Figure ~\ref{tab:oov-context} shows that \textit{\completion{}} has a lower 18.9\% OOV rate in the context tokens than the 25\% of \textit{\mercurial{}}, which helps explain why \textit{\completion{}} performs better on the \textit{\completion{}} test examples. The Transformer + BPE model has by far the lowest OOV rate for context tokens, which may contribute to the smaller relative decrease in accuracy we observe when training and evaluating on non-matching corpora.

\begin{figure}[]
\includegraphics[width=\columnwidth]{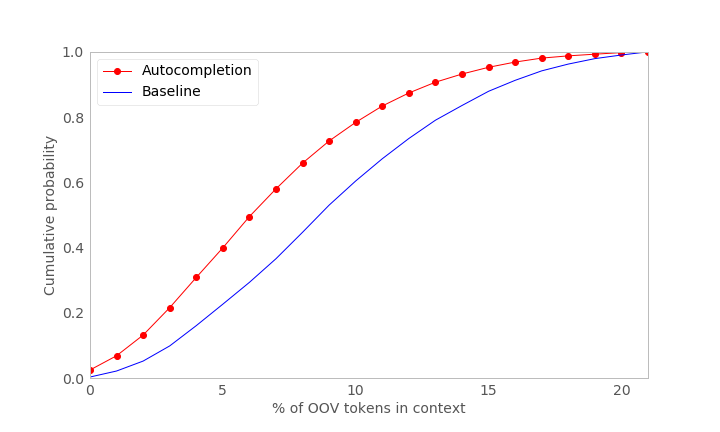}
\caption{Cumulative distribution function (CDF) of OOV tokens in the context vs top-1 accuracy for \textit{\mercurial{}} and \textit{\completion{}} Transformer models evaluated on the \textit{\completion{}} dataset.}
\label{fig:accuracy-oov-cdf}
\end{figure}

\paragraph{Length of accepted completion suggestions}

One noteworthy distribution difference between autocompletion tokens and random identifiers sampled from version control is that names used in autocompletion are averagely 12\% longer (14.31 characters versus 12.78). In fact, \cref{fig:accuracy-len-cdf} shows that 27.53\% of identifiers in version control are six characters or fewer compared to only 17.15\% of accepted completion suggestions. Intuitively, since one of the benefits of autocompletion is typing assistance, programmers save more time when using code completion for longer tokens. Additionally, longer lexemes may correspond to more complex and specialized APIs and library utilities which software developers will use code completion to discover and recall.

\paragraph{Target token kind}

Code completion can be used for local variables defined within the current scope or for non-local references such as a function imported through another module. While 35.34\% of identifiers in committed code are local variables, only 30.13\% of accepted code completions correspond to local variables. This distribution difference may cause \textit{\completion{}} to overestimate the likelihood of local variable usage in real code completion.

\paragraph{Recency bias}

The corpus of logged autocompletion events is collected by observing programmer activity over the course of ninety days at Facebook. One difference between examples in our study sourced from developer activity and examples sampled from source code in version control is that all of the real-world examples are guaranteed to be recently modified. Conversely, code sequences in committed code can be arbitrarily old. This difference leads to \emph{recency bias} in the real-world examples corpus since only 23.38\% of repository files were modified during the ninety day data collection period, indicating that the autocompletion events are drawn from a small, focused subset of the overall collection.

To investigate the impact of this bias, we train additional models on \textit{\edit{}} and a subset of \textit{\mercurial{}} limited to code sequences from files edited within the past ninety days. Both of these datasets share the recency bias of \textit{\completion{}}, but neither of these additional models exceed \textit{\mercurial{}} performance in offline evaluation. Whereas \textit{\completion{}} improves to 0.331 top-1 accuracy (from a 0.203 baseline) on real-world examples, \textit{\edit{}} and time-limited \textit{\mercurial{}} score 0.193 and 0.206 respectively (see \cref{tab:results-recency}). Since \textit{\completion{}} models are the only candidates to outperform \textit{\mercurial{}} amongst several models trained on datasets exhibiting recency bias, we conclude that their strength cannot be attributed to recency bias.

\begin{table}[]
	\caption{Offline evaluation results for Transformer models evaluated on the \textit{\completion{}} test corpus.}
	\def\arraystretch{1.20}
\small
\centering
\vspace{-1mm}
\begin{tabular}{c|c|c|c}
\toprule
Model & Training Corpus & Top-1 Accuracy & MRR \\
\midrule
\multirow{4}{*}{Transformer} & \mercurial{} & 0.203 & 0.267\\
\cline{2-4}
 & \completion{} & \textbf{0.331} & \textbf{0.408} \\
\cline{2-4}
 & \edit{} & 0.193 & 0.247 \\
\cline{2-4}
 & \mercurial{} (recent) & 0.206 & 0.270 \\
\bottomrule
\end{tabular}
\label{tab:results-recency}
\end{table}
\section{Threats to Validity}
\label{sec:threats}

\paragraph{Offline evaluation on one language} This study conducts offline evaluation using only one programming language to focus on the effects of training corpus selection on model accuracy. We train Transformer and n-gram models on combinations of three unique sources of training data and evaluate these models on each of the three corpora. While there are undoubtedly differences in token distributions and development practices between programming languages, we believe our findings apply to completion modeling in different languages as well. 

\paragraph{Data duplication} During data preprocessing, we do not dedupe examples between training and test datasets. Allamanis~\cite{allamanis2019adverse} explores this problem in depth. If code sequences are frequently copy-pasted or otherwise duplicated between train and test, the performance metrics we report may overestimate model strength. However, there is intuitive reason to believe that this effect is less pronounced in autocompletion examples as copy-paste events would not typically coincide with code completion.

\paragraph{Evaluation against other models} In this paper, we carry out our experiments using two models: n-grams and Transformers, both modeling code as a sequence of names. While there may be stronger models that exceed the performance of the ones we implement, our focus was not on the models, but rather on training corpus selection and its impact on model efficacy. Since the effects of selecting different training corpora hold across disparate sequence model types (n-gram and Transformer), we deduce that our findings apply generally across completion models.
\section{Related Work}
\label{sec:relatedwork}

Many studies at the intersection of software development tools and machine learning have investigated next code token prediction and its application to autocompletion. Earlier attempts at modeling software languages and completion ranking were based on n-gram language models\cite{hindle2012,nguyen2013,10.1145/3106237.3106290} or probabilistic models using the information from ASTs\cite{10.5555/3045390.3045699}. With the advancement of deep learning models, RNNs (and their variants)\cite{Li_2018,DBLP:journals/corr/abs-1903-05734} have shown promising improvements. More recently, Transformers have achieved state-of-the-art performance for software language modeling\cite{kim2020code,brockschmidt2019generative}. Galois (Radford et al., 2019) and TabNine™ are two additional code completion tools that employ the GPT-2 Transformer for next token prediction.

Furthermore, some studies have focused on the industry use case of autocompletion for IDE users within a company. Aye et al.\cite{aye2020sequence} shows that a pointer mixture network performs well on an open-source GitHub corpus as well as data internal to a large software company. Hellendoorn et al.\cite{hellendoorn-codefails} warns that accuracy achieved on synthetic benchmarks may not translate to real-world completion performance. Our work demonstrates how training on real-world developer activity data internal to a large software company can combat this challenge.
\section{Future Work}
\label{sec:future}

\paragraph{Copy mechanism}
Another popular approach in literature to handle OOV tokens is through a copy mechanism so that novel tokens from the context query can be predicted by reference (based on position). This strategy is especially helpful in software as local references are extremely common. Similar to subtoken encoding, incorporating copy mechanism may impact performance when moving from a synthetic benchmark to real-world examples.

\paragraph{Taking information after the cursor} 
In this paper, we focus on context tokens extracted from before (left and above) the cursor. However, it is uncommon for code to be written sequentially. We would like to explore further how code that follows the cursor (right and below) appears in the IDE and whether training models on developer activity creates opportunities for incorporating this additional signal.


\paragraph{Incorporating signal from the autocompletion prefix}
Typically IDEs will filter completion suggestions that do not match the prefix that a programmer has already typed. All of the models we train in this study ignore these prefix characters for the current token. This method matches approaches in existing literature and is appropriate when training on committed code as there is no way to know from looking at version control which characters a programmer will have typed when requesting code completion. However, training on real-world developer activity creates an opportunity to train on and learn from these prefixes, which are potentially high signal in code completion.
\section{Conclusion}
\label{sec:conclusion}

In this paper, we explored strategies for accurate and realistic evaluation of code completion models. In literature, most software language models are trained and evaluated on artificial examples from committed code in version control. However, when scoring these models on production logs from a deployed autocomplete tool, we observe a dramatic decrease in accuracy across n-gram and Transformer sequence models. This disparity indicates that traditional evaluation methods are unreliable for measuring improvements in real-world performance. 

We investigated several distribution differences between code sequences in the corpora studied to which we attribute the models' divergent ranking behavior. To combat the concept drift underlying this drop in accuracy, we constructed various training corpora (\textit{\mercurial{}}, \textit{\completion{}}, \textit{\edit{}}) and found that training on \textit{\completion{}} yields the best performance on the \textit{\completion{}} evaluation dataset. Furthermore, when models trained on the various corpora were deployed to rank completion suggestions in a live A/B test with thousands of programmers at Facebook, models trained on the \textit{\completion{}} dataset had the highest tool usage. 

\bibliographystyle{IEEEtran}
\bibliography{references.bib}

\end{document}